\documentclass[11pt,twoside]{article}
\usepackage{asp2010}

\resetcounters

\begin{document}

\title{Practices in Code Discoverability}
\author{Peter Teuben$^1$, Alice Allen$^2$, 
Robert J. Nemiroff$^3$, and Lior Shamir$^4$
\affil{$^1$Astronomy Department, University of Maryland}
\affil{$^2$Calverton, MD}
\affil{$^3$Michigan Technological University}
\affil{$^4$Lawrence Technological University}}

\begin{abstract}

Much of scientific progress now hinges on the reliability, 
falsifiability and reproducibility of computer source codes. 
Astrophysics in particular is a discipline that today leads other 
sciences in making useful scientific components freely available 
online, including data, abstracts, preprints, and fully published 
papers, yet even today many astrophysics source codes remain hidden 
from public view. We review the importance and 
history of source codes in astrophysics and previous efforts to develop 
ways in which information about astrophysics codes can be shared. 
We also discuss why some scientist coders resist sharing or publishing 
their codes, the reasons for and importance of overcoming this resistance, 
and alert the community to a reworking of one of the first attempts
for sharing codes,
the Astrophysics Source Code Library (ASCL). We discuss the implementation 
of the ASCL in an accompanying poster paper. 
We suggest that code could be given a similar level of referencing as 
data gets in repositories such as ADS.

\end{abstract}

\section{Introduction}

The importance of scientific codes has increased; indeed, this
importance is considered a fact of life \citep{Weiner2009} and is
continually being discussed in the literature\footnote{A larger
scoped paper on this topic is also under preparation by the authors}.
Many examples of public codes now exist
that have become industry standard software, such as
Sextractor, CLOUDY and GADGET to name a few.

In some fields (e.g., bioinformatics) journals include
software used to generate results with their articles or require it be
submitted. \cite{Gray2011} claim the astrophysics community is not
there yet, but scientists are encouraged to release their
codes, that their codes are good enough to release even if messy or
rough \citep{Barnes2010}. Scientists may see their codes, or
their research teams' codes, as proprietary and thus
refrain from publishing them.

Appropriate software is {\it “often equally important”}
\citep{Grosbol2010} as data are to research, and Weiner, et al. (2009) state
that {\it “useful public software packages … have enabled easily as much
science as yet another large telescope would have”}. Though the
NSF was specifically addressing
cyber-infrastructure with the statement that its strategic plan defines
“research infrastructure” as including investments in “experimental
tools”, we believe a case can be made for scientific codes
fitting within this strategic goal \citep{stewart2010}.

\section{A Brief History of Source Codes}

In the early years of computational astrophysics, several important
codes were developed but made available only to the social communities
surrounding their developers.  These social communities would
typically include close collaborators, graduate students, postdoctoral
fellows, and the graduate students of close collaborators.

An example of this is the "Wilson-Devinney" code that models eclipsing
binary star systems and their observable light curves.  A first
version of this code was written by Robert Wilson in or before 1971.
The Wilson-Devinney code has been upgraded and adapted numerous times.
Only recently has this code been made available via anon.ftp.

Another example is the {\it "Aarseth"} 
code that models gravitational N-body
interactions started around 1960 as one of the 
participants of the IAU 25 body problem ``contest''. 
Several versions of the Aarseth code are now made publicly available by 
its primary author, Sverre Aarseth through his web site.

\section{Previous Online Efforts}

With the rise of the internet, packages such as AIPS, IRAF, GIPSY
quickly became available to the community.

In the 1980s and 1990s, several prominent astrophysics codes were
released to the public over specific web sites.
Most of these were not associated with a specific
scientific paper.  Many of these sites still exist today.  The primary
way one found out about codes like these was through a mentor or
collaborator. In the general field of computing, websites sprang up
providing registration or repository service (e.g. freshmeat, sourceforge, 
github, google.code)

Two of the earliest code collections in astronomy were
AstroWEB and ASDS, but neither
are maintained anymore (though AstroWEB is still available at NRAO).

In 1999, Nemiroff and Wallin founded the online
Astrophysics Source Code Library (ASCL) to house codes of use
to the community, eventually resulting in a library of 37 codes,
all of which had been described in the literature and used to produce
research published in or submitted to refereed journals. 
The last code was 
added in 2002. The ASCL site also linked to other code libraries, most
of which no longer exist or have not been updated in years.

SkySoft  was
created in 2001 by C.Baffa, E.Giani, and A.Checcucci. This site
is intended to be a site which is community-supported,
accepting codes and comments from coders and code users. It also
features recent news on topics of interest to astronomer programmers,
such as notices of upcoming conferences and workshops for this
community. The majority of its code entries date from 2003, with some
additional codes from later years.

The Astroforge project was modeled after
the wildly successful SourceForge 
for open source software but focused on the needs of
astronomers (Remijan, Brunner, Tillery, \& Haider, 2003)
and existed for three years. 

We know of three other independent code information repositories,
though certainly there are many project groups and individuals who
pull together such information for their own work, team, or
subspecialty. One such subspecialty repository is AstrOmatic
for astronomical pipeline software;
another is the codes wiki for computational fluid dynamics

The Astro-Code Wiki 
created
by AstroSim - European Network for Computational Astrophysics
contains 54 codes and has been
updated recently. AstroSim was intended to be a five-year project “to
bring together European computational astrophysicists” running from
October, 2006 until September, 2011; its focus is on
comparison of codes for suitability for specific tasks.

Another repository called Astro-Sim houses about thirty codes, and
provides forums for discussion and links to other tools and
libraries. Similarly, AstroShare, discussed by \cite{Shortridge2009},
also houses about thirty codes and allows for discussion of topics
such as releasing software, social media, and middleware.

\section{To release or not to release source code}

Previous endeavors, including the first incarnation of the ASCL, had
not grown as codes have proliferated. Indeed, some scientists are not
in the habit of, are reluctant to, or openly resist making their codes
available to non-collaborators. A look at popular and academic
literature, our correspondence and conversations with scientist
coders, informal surveys, and our experiences demonstrate the variety
of reasons some scientists have for not releasing codes.

\noindent{\bf Intellectual property issues:}
The workplace or granting institutions may place restrictions
on sharing code.

\noindent{\bf Codes reflect the reality of their creation:} 
Code is often "quick and dirty"; because it is messy, a coder may be
reluctant to release it. Codes also can have a narrow focus, and the
author(s) doesn’t seem it suitable for anything else.

\noindent{\bf Releasing codes is not standard practice, useful to
  one’s career, necessary, nor desired:} it is a fact of life that
coding does not get you brownie points.

\noindent{\bf Releasing codes places demands on the coder, and
  released codes may be examined too closely and used
  inappropriately:} programmers may be worried there might be bugs in
their code (Barnes, 2010). This is again a problem of the lack of time
to check code results to some subjective level of increasingly greater
tolerance when this time could be used to write more papers and again
advance in the "publish of perish" dilemma.

\subsection{Why codes should be released}

Despite these arguments and absent any national security concerns, we
believe it is incumbent upon scientists to release their codes. If a
code does the job it was designed to do, it does not matter that the
code may be messy, undocumented, or cobbled together and inelegant; as
a tool used to produce results, the code should be available for
examination and study, just as any research protocol is.

We are not alone in this belief. Timmer\footnote{\url{http://arstechnica.com/science/news/2010/01/keeping-computers-from-ending-sciences-reproducibility.ars}}
laments that the
reliance on computational methods in the sciences has scientists
giving up “{\it on a key component of the scientific method:
reproducibility.}” A lack of transparency and reproducibility
“{\it undermines public confidence in science as well as slowing scientific
progress, engendering a credibility crisis,}” according to 
Stodden\footnote{\url{http://www.stanford.edu/~vcs/AAAS2011/}}
who is working to develop the Reproducible Research Standard.

The NSF has made a
recommendation that reproducibility should be promoted, and states
that “{\it data and software used in the development of a scientific
publication should be escrowed or archived where they can be examined
and re-verified when needed}” \citep{stewart2010}.

\section{ASCL}

We have implemented a new way to provide a large set
of peer-reviewed described codes from the community in an easily
accessible place. The details are described
in our accompanying paper \citep{P003_adassxxi}. An additional
outcome of such a repository could be a referencing database
for astrophysics code, similar to the newly established one for
data that has been added to the ADS, cf. discussions
during a BoF session \citep{B1_adassxxi}.

{\bf NOTE ADDED IN PROOF:}  ASCL codes are now 
incoorporated into ADS.


\bibliography{O27}

\end{document}